%% file: main.tex
  \documentclass[journal]{IEEEtran}

  \usepackage[absolute,overlay]{textpos}

\usepackage{balance}
\usepackage{graphicx}
\usepackage [noadjust]{cite} 
\usepackage{bm}  
\usepackage{amsmath}
\usepackage{amssymb}
\usepackage{enumerate}
\usepackage{stfloats}
\usepackage{cases}
\usepackage{epstopdf}
\usepackage{soul,color} 
\usepackage{hyperref}
\usepackage[percent]{overpic}
\usepackage{tabularx}
\usepackage[table]{xcolor}
\usepackage{multirow}
\usepackage{booktabs}  
\usepackage{tabu} 
\usepackage{amsthm} %
\usepackage{xr}

\usepackage{algorithm}
\usepackage{algpseudocode}

\newcommand{\cb}{\color{black}}

\theoremstyle{remark}   
 
\begin{document}
	\raggedbottom
	\allowdisplaybreaks
    \title{
    Beam Squint Mitigation in Wideband Hybrid Beamformers: Full-TTD, Sparse-TTD, or Non-TTD?

  \thanks{
  This research is supported by Research Council of Finland via 6G Flagship Programme under Grant 369116, Research Council of Finland via Profi6 Project under Grant 336449, and Business Finland via  6GBridge - Local 6G Project under Grant 8002/31/2022.
    }
    }
	\author{Mehdi~Monemi, \textit{Member, ~IEEE}, Mohammad Amir Fallah, Mehdi~Rasti, \textit{Senior~Member,~IEEE}, Omid~Yazdani, Onel~L.~A.~López,~\textit{Senior~Member,~IEEE}, Matti~Latva-aho, \textit{Fellow,~IEEE}
 \thanks{
  M. Monemi, M.~Rasti O.~L.~A.~López, and M. Latva-aho are with Centre
for Wireless Communications (CWC), University of Oulu, 90570 Oulu, Finland (emails: mehdi.monemi@oulu.fi, 
mehdi.rasti@oulu.fi, onel.alcarazlopez@oulu.fi, matti.latva-aho@oulu.fi).
}
\thanks{
  M. Amir Fallah is with Department of Engineering, Payame Noor University (PNU), Tehran, Iran (email: mfallah@pnu.ac.ir)
  }
\thanks{
O. Yazdani is with the Department of Electrical
Engineering, Yazd University, Yazd, Iran (email: omid.yazdani@stu.yazd.ac.ir)
}
}

	\maketitle
	\begin{abstract}

Beam squint poses a fundamental challenge in wideband hybrid beamforming, particularly for mmWave and THz systems that demand both ultra-wide bandwidth and high directional beams. While conventional phase shifter-based beamformers may offer partial mitigation, True Time Delay (TTD) units provide a fundamentally more effective solution by enabling frequency-independent beam steering.  However, the high cost of TTD units has recently driven much interest in Sparse-TTD architectures, which combine a limited number of TTDs with a higher number of conventional PSs to balance performance and cost. 
This paper provides a critical examination of beam squint mitigation strategies in wideband hybrid beamformers, comparing Full-TTD, Sparse-TTD, and Non-TTD architectures. We analyze recent Non-TTD approaches, specifically the scheme leveraging the wideband beam gain (WBBG) concept, evaluating their performance and cost characteristics against TTD-based solutions. A key focus is placed on the practical limitations of Sparse-TTD architectures, particularly the often-overlooked requirement for wideband PSs operating alongside TTDs, which can significantly impact performance and implementation cost in real-world scenarios, especially for ultra-wideband applications. Finally, we conduct a cost-performance analysis to examine the trade-offs inherent in each architecture and provide guidance on selecting the most suitable hybrid beamforming structure for various fractional bandwidth regimes.

	\end{abstract}
	
	
\thispagestyle{empty}

\section{Introduction}

\begin{textblock*}{11cm}(9.4cm,1cm)  
   Accepted for publication in IEEE Wireless Communications Magazine
\end{textblock*}

Efficient beamforming is essential for enabling 6G technologies that exploit mmWave and THz bands to deliver ultra-high data rates and dense connectivity.In phased array systems, analog beamforming adjusts signal phases of antenna elements utilizing a single RF chain, while digital beamforming offers fine-grained amplitude and phase control at the cost of one RF chain per antenna element. To balance performance and hardware complexity, hybrid beamforming combines both approaches by connecting fewer RF chains to large antenna arrays via analog phase shifters (PSs). This allows flexible beam control with reduced cost and power consumption.
In contrast to conventional beamforming methods, which often rely on narrowband assumptions, contemporary applications increasingly demand wideband and ultra-wideband beamforming to support ultra-high data rates as well as robust communications. For instance, in an ultra-wideband multi-carrier transmitter, frequency diversity enhances robustness by allowing the system to maintain performance despite deep fading or interference on specific carriers, as unaffected carriers can ensure reliable operation \cite{monemi2025higher}. This is particularly vital for frequency-selective fading channels, driving significant research into advanced wideband beamforming techniques that underpin the evolution of resilient, high-capacity 6G systems. 
Despite its advantages, wideband beamforming introduces a critical challenge known as beam squint. Beam squint arises when fixed phase shifts optimized for a target frequency $f_0$, lead to frequency-dependent beam direction change for frequency $f\neq f_0$. For the simple case of a uniform linear array (ULA) transmitter, the main lobe of the radiated beam at frequency $f$ is deviated by $\Delta \theta =\sin^{-1}(\frac{f_0}{f}\sin(\theta_0))-\theta_0$, where $\theta_0$ corresponds to the intended beam angle at frequency $f_0$ with respect to the boresight of the antenna. This frequency-dependent beam squint degrades beamforming gain and overall system performance, particularly in ultra-wideband systems. 

To mitigate beam squint in wideband radar and communication systems, several techniques have been proposed. The first solution scheme focuses on implementing signal processing and beamforming techniques implemented using conventional analog-domain PSs. For instance, in the context of  Multi-Input Multi-Output (MIMO) wideband radar, the work in \cite{8713914} addresses beam pattern design under a space-frequency nulling constraint, aiming to suppress frequency-dependent interference caused by beam squint. In \cite{10378951}, the authors propose a beamforming scheme for wideband radars that synthesizes frequency-invariant beam patterns via constrained waveform design, incorporating nulling and peak-to-average power ratio control in the space-frequency domain. More recently, \cite{10739944} introduced a hybrid beamforming approach that broadens the beam in the angular domain to achieve max-min fair beam gain across spatially squinted beams. 

\begin{figure*}
    \centering
    \includegraphics[width=1\linewidth]{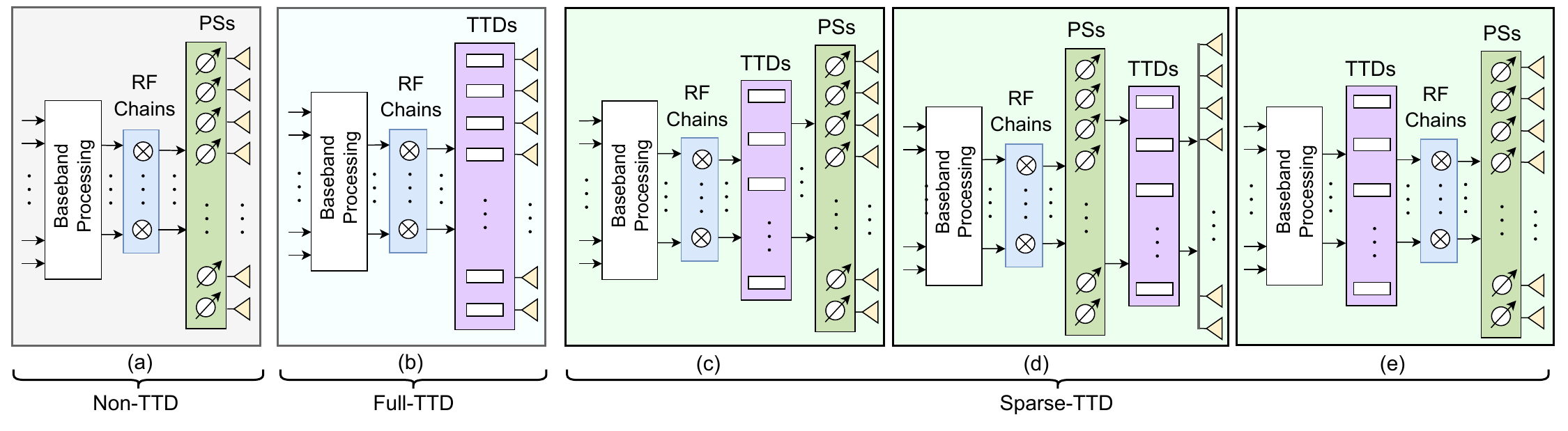}
    \vspace{-10pt}
    \caption{
    Wideband hybrid beamformer leveraging (a) conventional phase-shifters (PSs) without deploying TTDs (Non-TTD), (b) one TTD per antenna element (Full-TTD), and (c): sparse TTDs together with a block of PSs (Sparse-TTD)}
    \label{fig:structures}
\end{figure*}

Wideband beamformers using conventional PSs can partially mitigate beam squint, as they cannot provide an optimal solution due to the group delay associated with such components.
In this regard, the second approach aims to more fundamentally mitigate beam squint by replacing phase shifters with true time delays (TTDs).  Unlike phase shifters that introduce frequency-dependent phase shifts, TTDs apply frequency-independent delays, ensuring consistent beam direction across a wide bandwidth. 
In general, beam squint mitigation through Non-TTD-based hybrid structures follows the general block diagram of a conventional hybrid beamformer illustrated in Fig.~\ref{fig:structures}-a ~\cite{10739944}
. This structure which employs only PSs, suffer from performance degradation when operating over large fractional bandwidths. In this regard, the second approach aims to more fundamentally mitigate beam squint by replacing PSs with true time delays (TTDs). Unlike PSs that introduce frequency-dependent phase shifts, TTDs apply frequency-independent delays, ensuring consistent beam direction across a wide bandwidth.
     \textit{Full-TTD architectures} where each antenna element is equipped with a dedicated TTD as shown in Fig.~\ref{fig:structures}-b~\cite{9852650}, are known to optimally mitigate beam squint by providing 
    frequency-independent delays for each antenna element. However, practical challenges such as 
    high hardware cost makes this design inefficient for ultra-wideband applications, especially in mmWave and THz frequencies. To reduce this overhead, instead of providing all antennas with dedicated TTDs, a lower number of TTD ({\it Sparse-TTD}) can be leveraged using one of the structures illustrated in Figs.~\ref{fig:structures}-c–e. 
    The structure in Fig.~\ref{fig:structures}-c places the TTD block after the RF chains and before the PSs, which allows efficient coarse delay compensation followed by per-antenna phase adjustment~\cite{10225726,10440417}.
    Alternatively, in the architecture in Fig.~\ref{fig:structures}-d utilized in some works such as  ~\cite{10179244}, TTDs are placed after PSs, thereby providing highly accurate delay adjustment in the final stage. This architecture, however requires a more complex structure for interconnecting the TTD block and antenna elements. 
    Finally, Fig.~\ref{fig:structures}-e places the TTDs before the RF chains ~\cite{10381610}, allowing their operation at low-power signal levels. This reduces insertion loss and makes it feasible to employ low-power, cost-effective TTDs.

Given the relatively high cost of TTDs compared to PSs, a Sparse-TTD configuration aims to reduce the hardware cost of a Full-TTD wideband hybrid beamformer. However, this approach presents several notable drawbacks. First, it does not completely eliminate beam squint, leaving residual misalignment across some frequency bands, since full squint mitigation requires equipping each antenna with a dedicated TTD. Second, the joint optimization of 
precoders, TTD delays, and PSs, introduces considerable computational complexity, as the design must jointly optimize all these to achieve near-optimal performance. This is commonly followed by a series of decomposition and approximation techniques that can further degrade the performance \cite{10440417,10381610}. Finally, and most critically, a challenge often overlooked in the literature is the requirement for \textbf{wideband PSs} to operate alongside TTDs. This necessity introduces additional implementation concerns, which will be examined in detail in this work. 

\begin{figure}
    \centering
    \includegraphics[width=1\linewidth]{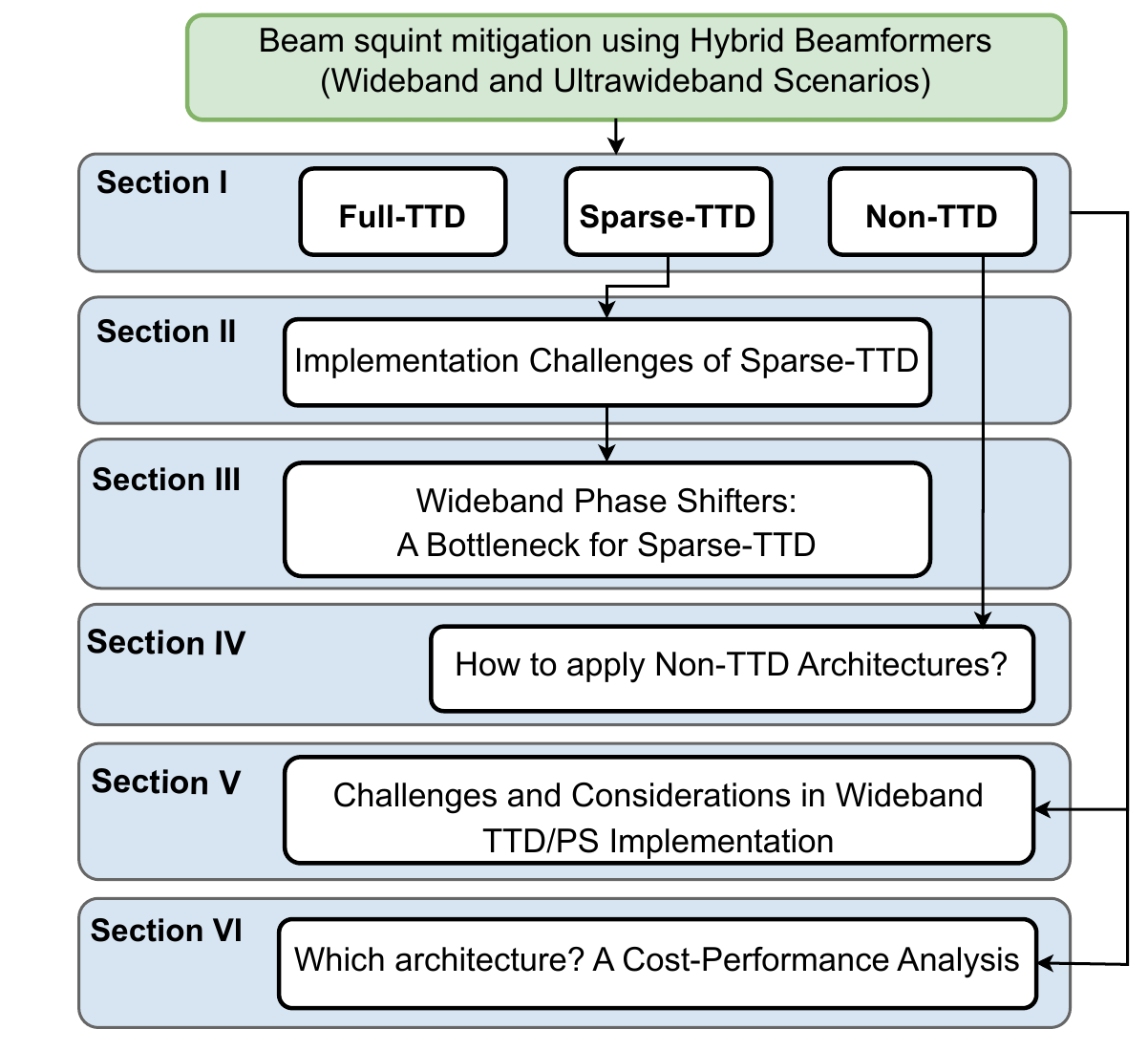}
    \vspace{-10pt}
    \caption{The hierarchical structure of the article.}
    \label{fig:structure}
\end{figure}

 Following the discussions provided so far relating to inherent differences of the three aforementioned architectures, the main contributions of this paper associated with the organization visualized in Fig.~\ref{fig:structure} are summarized as follows:
\begin{itemize}
    \item  While Sparse-TTD wideband hybrid beamformers are commonly cited as a cost-effective, high-performance solution for mitigating beam squint, in this paper we critically examine the practical limitations. In particular, beyond their inherently high computational complexity, we highlight the often-overlooked requirement for wideband PSs and discuss how this constraint can significantly affect performance and implementation cost in real-world scenarios.  These are explored in Sections \ref{sec:345kjsdg} and \ref{sec:345kjsdg2} as illustrated in Fig. \ref{fig:structure}.

    \item We analyze recent non-TTD-based techniques for beam squint mitigation implemented through wideband hybrid beamformers. A comparative evaluation is provided to examine the advantages and drawbacks of these approaches relative to Sparse-TTD and Full-TTD architectures. This will be covered in Section \ref{sec:Sfhus52}. 

   \item We perform a cost-performance analysis of Non-TTD, Sparse-TTD, and Full-TTD hybrid beamforming architectures as a function of fractional bandwidth. We analyze how the performance and cost-efficiency of Non-TTD and Sparse-TTD beamformers degrade more significantly with increasing fractional bandwidth compared to the Full-TTD configuration. Based on the analysis conducted, we recommend optimal beamformer structures for various bandwidth regimes, considering both performance and cost considerations.

\end{itemize}

\section{Sparse-TTD Beamforming: Implementation Challenges}
\label{sec:345kjsdg}

While sparse-TTD hybrid beamformers are often regarded as high-performance and cost-effective solutions for beam squint mitigation, their practical deployment faces several challenges. To explore these limitations, we first present the corresponding beamformer optimization formulation. The problem for a Sparse-TTD-based multi-carrier wideband beamformer can be expressed as follows:
 \begin{align}
 \label{eq:678327ss}
     \max_{\mathbf{D}_m, \mathbf{T},\mathbf{A}}\ &  Q(\mathbf{D}_m,\mathbf{T},\mathbf{A})
     \notag
     \\
     \mathrm{s.t.\ } & G_i(\mathbf{D}_m,\mathbf{T},\mathbf{A})<0,\forall i,
 \end{align}
where $\mathbf{D}_m$ represents the digital beamformer corresponding to each carrier $m$, $\mathbf{T}$ is the vector encompassing the tunable delays of the TTDs ($0\leq T_n\leq T^{\mathrm{max}}$ for each TTD $n$), and finally $\mathbf{A}$ is the analog beamformer, typically realized through conventional PSs. The objective function $Q$ captures the desired QoS criterion, such as the sum spectral efficiency (sum-SE) across all carriers, and finally $G_i,\forall i$ correspond to feasibility constraints, such as those imposed by transmit power limitations. While widely adopted for beam squint mitigation in recent works in the literature, the Sparse-TTD structure presents several challenges that are often overlooked in addressing the associated optimization problem. These limitations might significantly restrict its applicability across various practical scenarios. In the following, we outline the key practical challenges associated with this structure:
    
    $\bullet$
    One key limitation lies in the inherent complexity of the beamforming optimization problem. The Sparse-TTD architecture requires the joint optimization of the interdependent parameter sets: the digital precoders $\mathbf{D}_m$ for each subcarrier $m$, the tunable delay vector $\mathbf{T}$ for the TTDs, and the analog beamformer $\mathbf{A}$. This multi-component formulation introduces significant computational complexity due to the high dimensionality and non-convex nature of the problem, particularly in large-scale systems with many carriers and antenna elements. While various approximation and convexification techniques have been proposed in the literature to address this challenge, they often result in errors due to approximations that further diverge from the ideal performance of a Full-TTD design. 

    $\bullet$
   The globally optimal solution of such highly complex non-convex problems requires greedy searches over the domain of ${\mathbf{A},\mathbf{T},\mathbf{D}_m,\forall m}$, which entails exploring extremely large codebooks and is generally intractable. The strong coupling among the multi-dimensional decision variables instead motivates decomposition strategies, where multi-level alternating optimization (AO) is commonly adopted in related works (e.g., \cite{10440417,10381610}). In this approach, the original problem is decomposed into three subproblems, each optimizing one variable set while keeping the others fixed, and the overall solution is obtained iteratively. While this framework improves tractability, it still suffers from high computational cost and potential convergence issues. In particular, most existing studies do not provide convergence guarantees or complexity characterizations. Moreover, even if  convergence can be ensured, the process is generally slow because of the large dimensionality of the three decision-variable sets and the strong coupling among them.
     $\bullet$
     Beyond algorithmic challenges, hardware limitations for addressing the solution to the formulated optimization problem significantly influence the performance of Sparse-TTD-based hybrid beamformers. One of the most critical challenges generally overlooked in the literature is the requirement of frequency-independent wideband PSs corresponding to the optimal decision variables of $\mathbf{A}$. In more detail, as opposed to the digital precoders $\mathbf{D}_m$ where a dedicated precoder vector is assigned for each carrier $m$, the optimal value of $\mathbf{A}$ is frequency independent, which should remain fixed for all carriers across the whole bandwidth. This is a real challenging issue that can practically influence the performance and cost of the system, as elaborated in the next section.

\section{Wideband Phase Shifters: A Bottleneck for Sparse-TTD Wideband Beamformers}
\label{sec:345kjsdg2}

As previously discussed, the optimal analog beamformer $\mathbf{A}^*$, corresponding to the solution of the optimization problem in \eqref{eq:678327ss}, requires the realization of frequency-independent phase shifts that remain constant across all subcarriers, irrespective of the carrier index $m$. As the bandwidth increases, achieving such frequency-invariant phase shifts becomes increasingly challenging and costly. Depending on the fractional bandwidth, operating frequency band, and required phase resolution, various techniques may be employed to realize an approximately constant phase shift in the available wideband spectrum. These techniques generally fall into three categories: two non-TTD-based approaches and one TTD-based approach, as detailed below. The first method uses a set of wideband fixed PS components to construct a dynamically tunable wide-range PS, implemented using a hybrid of multiple passive architectures.
A representative design is presented in \cite{10023445}, where a 6--18\,GHz 6-bit wideband PS is realized using several circuit topologies: a switched transmission line for the smallest bit ($5.625^\circ$), high-pass/low-pass filters for intermediate bits ($11.25^\circ$, $22.5^\circ$, $45^\circ$), a magnetically coupled all-pass network for the $90^\circ$ bit, and a Lange-based structure for the $180^\circ$ bit. The second method uses conventional narrow-band phase-shift cells, which exhibit frequency-dependent phase drift, and augments them with broadband equalizers or negative-group-delay (NGD) networks \cite{meng2022broadband} leveraging circuit components having controllable impedance. These equalizer/NGD sections are designed to have a complementary group-delay-versus-frequency characteristic that compensates for the phase error across the band of interest. Finally, the third method builds upon the previous approach by replacing the narrowband phase-shift cells with wideband TTDs followed by NGD circuits~\cite{ravelo2023dual}, such that the positive group delay introduced by the TTD is compensated by the NGD.

As the fractional bandwidth increases, the cost and complexity of implementing wideband PSs escalate significantly \cite{kebe2024survey}, often making the overall cost of beamformers employing the first two categories of PSs comparable to, or even exceeding, that of TTD-based counterparts. Furthermore, since the third method incorporates TTD units within the PS structure itself, applying this approach to Sparse-TTD hybrid beamformers is fundamentally discouraged, as it increases the number of required TTDs, contradicting the primary motivation behind Sparse-TTD architectures, namely, reducing the TTD count to minimize cost and complexity. Considering the discussions presented, a key insight can be drawn regarding the practicality of Sparse-TTD beamformers:

\textit{The requirement for wideband PSs poses a critical challenge to the cost-effectiveness and performance of Sparse-TTD hybrid beamformers in wideband and ultra-wideband scenarios, potentially offsetting their advantages over Full-TTD architectures.}

\begin{figure}
    \centering
    \includegraphics[width=1\linewidth]{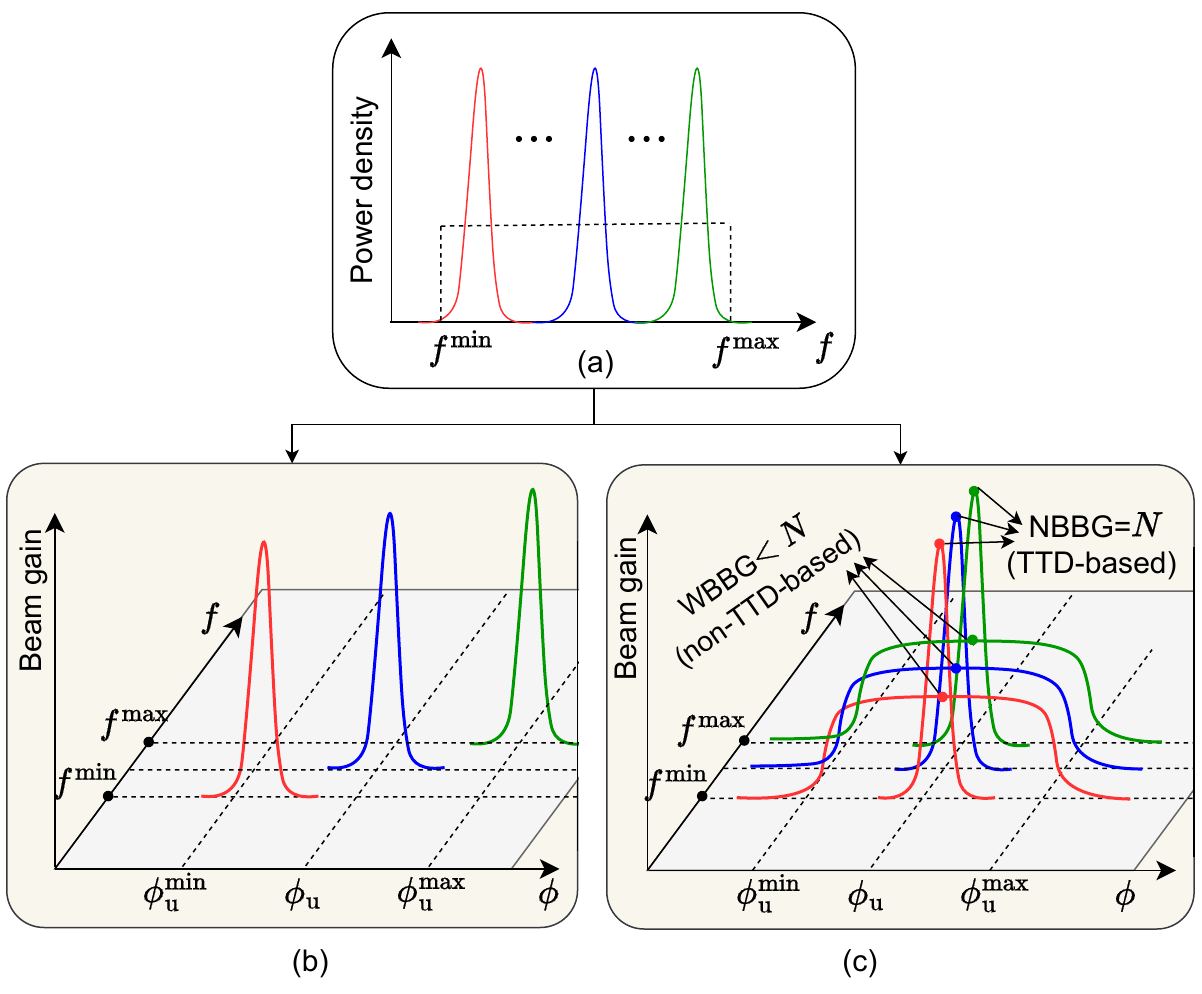}
    \vspace{-15pt}
    \caption{(a) Power density versus carrier frequency in a multi carrier wideband system. (b) Representation of beam gain in the angular and spectral domains when no beam squint mitigation is utilized. (c) Comparison of beam squint removal leveraging Full-TTD MRT vs. Non-TTD WBBG mechanisms.}
    \label{fig:WBBG}
\end{figure}
\section{Combating Beam Squint in Non-TTD-based Hybrid Beamformers}
\label{sec:Sfhus52}
The high cost and implementation complexity of TTD units have motivated the development of alternative beam squint mitigation strategies for hybrid beamformers that avoid the use of TTDs. For instance, the authors of \cite{10739944} recently proposed a Non-TTD-based wideband hybrid beamforming scheme based on a new metric, the \textit{wideband beam gain} (WBBG), which leverages conventional PSs. However, the performance of this approach has not been compared or analyzed against TTD-based solutions. This section outlines the core concept of the Non-TTD-WBBG method and discusses its strengths and limitations relative to TTD-based architectures.

Consider a transmitter equipped with $N$ antenna elements transmitting multi-carrier signals over a wide frequency band spanning from $f^{\mathrm{min}}$ to $f^{\mathrm{max}}$, as illustrated in Fig.~\ref{fig:WBBG}-a, where the intended user is located at the angular position $\phi_{\mathrm{u}}$ and is expected to receive signals from all carriers. When conventional narrowband maximum-ratio transmission (MRT) beamforming is employed using frequency-independent (wideband) PSs, the signals at different carrier frequencies are steered toward different directions, deviating from $\phi_{\mathrm{u}}$. This frequency-dependent deviation results in an angular spread ranging from $\phi_{\mathrm{u}}^{\mathrm{min}}$ to $\phi_{\mathrm{u}}^{\mathrm{max}}$, corresponding to $f^{\mathrm{min}}$ and $f^{\mathrm{max}}$, respectively, as depicted in Fig.~\ref{fig:WBBG}-b. This beam squint effect causes each carrier’s beam to be radiated towards a distinct angle, often placing the user outside the main lobe of many carriers. 
A Full-TTD architecture can counteract beam squint by aligning all frequency components toward $\phi_{\mathrm{u}}$, thereby achieving the full array gain of $N$ for every carrier, as illustrated in the figure. In contrast, the Non-TTD-WBBG approach proposed in~\cite{10739944} mitigates beam squint using conventional PSs. This method optimizes the PS values to maximize the minimum beam gain across all carriers within the angular interval $[\phi_{\mathrm{u}}^{\mathrm{min}}, \phi_{\mathrm{u}}^{\mathrm{max}}]$, resulting in a relatively flat wideband beam pattern, as shown in Fig.~\ref{fig:WBBG}-c. Unlike conventional MRT, the WBBG technique provides approximately uniform beam gain across all carriers, thus ensuring max-min fairness under beam squint.

However, this approach introduces several practical challenges and trade-offs, as outlined below:

\begin{itemize}
    \item Broadening the beam pattern to cover the angular squint range reduces the beam gain compared to the ideal narrowband array gain of $N$, which can be achieved using TTDs, as illustrated in Fig.~\ref{fig:WBBG}-c. This degradation becomes more pronounced with increasing fractional bandwidths, as a wider angular squint range results in a lower wideband beam gain, thereby leading to weaker overall performance.

    \item   Optimizing the max-min fairness criterion for large-scale antenna arrays across a wide range of carrier frequencies introduces substantial computational complexity, particularly in the hybrid mode for jointly designing the digital precoders and analog beamformers. This complexity is generally much higher than addressing hybrid beamforming involving only max or min objectives. In practice, the problem is often addressed through a combination of convex approximations and iterative algorithms such as bisection methods \cite{10739944}. However, these approximations can lead to a notable performance gaps.

    \item Similar to the Sparse-TTD architecture discussed earlier, the Non-TTD-WBBG scheme requires {\it wideband PSs}, the effective deployment of which poses significant challenges in terms of cost and efficiency in wideband and ultra-wideband communication scenarios. 
\end{itemize}
In summary, Non-TTD-WBBG provides a viable alternative for moderate bandwidths but experiences performance degradation in wideband and ultra-wideband scenarios. Moreover, its cost-effectiveness is uncertain due to the complexities of implementing wideband PSs, as discussed before.

\section{Challenges and Considerations in Wideband TTD/PS Implementation}

This section examines the practical implementation challenges of wideband PSs and TTDs across the three aforementioned beamforming architectures. Technical specifications and key limitations for several commercial and lab-fabricated implementations, ranging from lower frequency bands to the sub-THz regime, are summarized in Table~\ref{tab:ps_ttd_comparison}. 

\begin{table*}[htbp]
\centering
\caption{Representative Wideband PSs and  TTDs and Key Associated Loss/Error Values}
\label{tab:ps_ttd_comparison}
\small
\renewcommand{\arraystretch}{1.25}
\begin{tabular}{p{7cm}p{1.5cm} p{2cm}p{1.5cm} p{2.4cm}p{1.5cm}}
\hline
\textbf{Device} & \textbf{Resolution} & \textbf{Freq. Range}  & \textbf{Time Delay Range}  & \textbf{Max Phase/Delay Error} & \textbf{Insertion Loss} \\
\hline

\multicolumn{6}{l}{\textbf{Commercial Digitally Controlled Wideband PSs}} \\
\hline
PS-100M20G-10B-SFF-MIL-SM (Quantic PMI)
& 10 Bit  
& 0.1--20 GHz
& N/A
& $\leq \pm6^\circ$
& 0-13 dB \\

TGP2105-SM (Qorvo)
& 6-bit 
& 6--18 GHz
& N/A
& $\leq \pm8^\circ$
& 6-10 dB \\

\hline
\multicolumn{6}{l}{\textbf{Commercial Digitally Controlled Wideband TTDs}} \\
\hline
ADAR4002 TDU (Analog Devices)
& 7 Bit
& 0.5--19 GHz 
& 254 ps
& $\leq \ 2$ ps
& 1-20 dB 
\\
MM7005T (Miller MMIC)
& 4 Bit
& 1--12 GHz
& 390 ps
& $\leq \ 10$ ps
& 6-9 dB \\

\hline
\multicolumn{6}{l}{\textbf{Commercial Sub-THz Digitally Controlled Wideband PS and TTD}} \\
\hline
529 Series (Mi-Wave)
& 4 Bit
& 75--110 GHz
& N/A
& $\leq \pm4^\circ$
& 0-3 dB \\

 https://doi.org/10.23919/EuMC48046.2021.9338128
& continuous
& 53--120 GHz
& N/A
& N/A
& 0-3 dB \\

\hline
\end{tabular}
\end{table*}

\subsection{Wideband PSs}
From a hardware realization standpoint, wideband and ultra-wideband beamforming architectures are fundamentally constrained by implementation-level impairments in digitally controlled PSs. Unlike idealized frequency-flat models, commercial wideband PSs exhibit measurable frequency-dependent phase distortion and frequency-dependent insertion loss across their operating bandwidth. For example, as seen in Table~\ref{tab:ps_ttd_comparison}, the Qorvo TGP2105-SM operating over 6--18~GHz reports a frequency-dependent phase error of $[-8^\circ, +8^\circ]$ and an insertion loss ranging from 6 to 10~dB across the whole bandwidth. These distortions stem from dispersion in transmission-line networks, reactive elements employed in switched filter cells, and parasitic effects introduced by semiconductor switching networks. For small fractional bandwidths, these impairments remain moderate, rendering PS-based architectures compact and cost-effective with negligible amplitude/phase distortion.
\subsection{Wideband TTDs}
In contrast to the frequency-dependent phase distortion of wideband PSs, commercial wideband TTDs 
aim to enforce  frequency-invariant delay and exhibit less severe frequency dependent insertion loss, but introduce a different set of hardware challenges; they typically exhibit {\it delay-dependent insertion loss} 
and nonzero delay error across the operating band due to switching nonidealities and dispersion in delay lines. As shown in Table~\ref{tab:ps_ttd_comparison}, representative devices such as the ADAR4002 TDU and MM7005T demonstrate delay errors on the order of a few picoseconds over wide bandwidths. Note that although commercial TTDs introduce delay-dependent insertion loss and signal distortions, compensating for these impairments is generally less challenging than mitigating the frequency-dependent distortions inherent to wideband PSs. Since TTDs required delays are initially computed by the beamformer and applied to each TTD, amplifiers with gains adjusted proportionally to the applied calculated delays can be readily employed to alleviate associated loss and distortion.


\subsection{Other Impairments and Sub-THz Considerations}

Beyond standard frequency-dependent phase, delay, and amplitude distortions, wideband {\it electrical} PSs and TTDs suffer from electrically large parasitics, high dielectric losses, and extreme sensitivity to fabrication tolerances. These factors are further compounded by thermal fluctuations, which modify material properties and physical dimensions as well as device bias points, inducing time-varying phase errors and cumulative delay drift. While complex multi-section electrical architectures can attempt to equalize these responses, they incur excessive power consumption and calibration overhead. These implementation challenges become significantly more pronounced at sub-THz frequencies, where ohmic and dielectric losses scale aggressively, parasitic reactances become dominant, and structural fabrication margins represent a substantial fraction of the operating wavelength. {\it Photonic} architectures provide a robust alternative to mitigate these electrical limitations. At sub-THz frequencies, photonic wideband TTDs are significantly simpler to implement than  PS counterparts because frequency-independent delays arise naturally from optical path lengths or dispersion. In contrast, photonic PSs require stringent wavelength, thermal, and fabrication tolerances to achieve stable wideband operation. The final two rows of Table~\ref{tab:ps_ttd_comparison} present sub-THz wideband commercial electrical wideband PS and photonic TTD, respectively.

\cb

\section{Optimal Wideband Hybrid Beamformers: A Cost-Performance Trade-off Analysis}
\begin{figure}
    \centering
    \vspace{-10pt}
\includegraphics[width=1\linewidth]{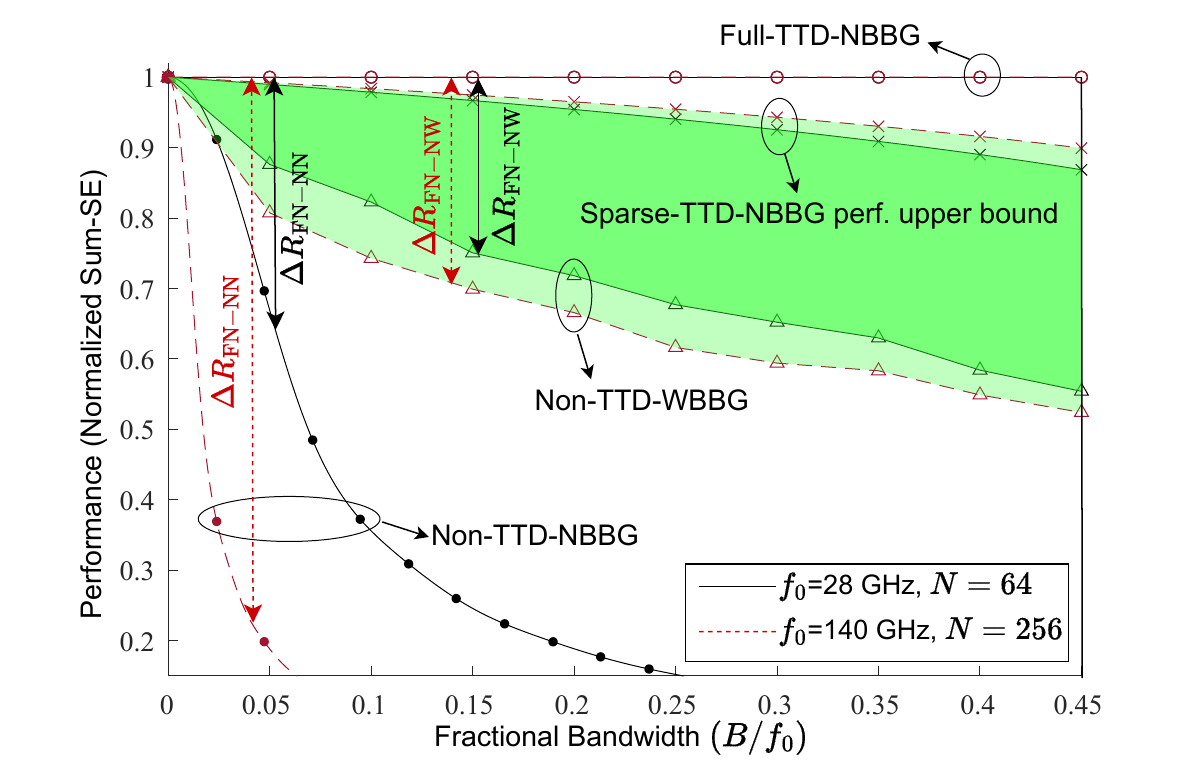}
\label{fig:pc}
{ \footnotesize (a)}
\includegraphics[width=1\linewidth]{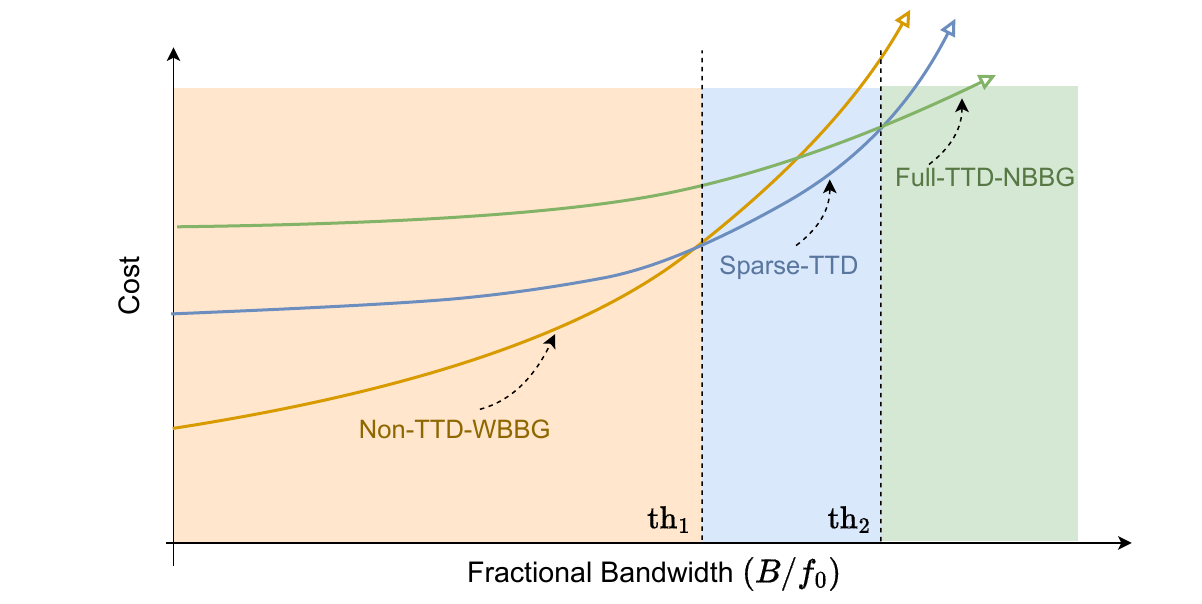}
\label{fig:c}
{\footnotesize (b)}
    \caption{{\bf(a)}: Performance versus fractional bandwidth for various beamformers and two setups including  mmWave with $f_0=28$ GHz and $N=64$ antennas, and sub-THz with $f_0=140$ GHz and $N=256$ antennas. We consider $M = 16$, $\tilde{\gamma}_0 = 0$~dB, and $\theta_{\mathrm{u}}=60^\circ$. 
    Depending on the design parameters, the Sparse-TTD beamformer's achieved performance lies at some point within the upper and lower bounds shown as the green-shaded region.
    \\
     {\bf (b)}: Cost trend versus  operating fractional bandwidth. 
     The regions where Non-TTD-WBBG, Sparse-TTD-NBBG, and Full-TTD-NBBG beamformers are more cost-effective are highlighted in orange, blue, and green, respectively.}
     \label{fig:pc-c}
\end{figure}


As discussed earlier, the fractional bandwidth plays a pivotal role in shaping both the performance and cost of wideband hybrid beamformers, making the selection of an optimal architecture a non-trivial design challenge,   In this section, we present a cost-performance analysis using simplified system models to provide insights into this trade-off. 
Fig.~\ref{fig:pc-c}-a compares the performance of several hybrid beamforming architectures: Non-TTD wideband beam-gain (Non-TTD-WBBG), Non-TTD narrowband beam-gain (Non-TTD-NBBG), Full-TTD-NBBG, and Sparse-TTD-NBBG, all evaluated as functions of the fractional bandwidth $B/f_0$,  where $B$ is the total bandwidth, and $f_0$ represents the central carrier frequency. Unlike the Non-TTD-WBBG, which explicitly broadens the beam across the entire spectrum, the other architectures optimize performance without explicitly considering inter-carrier frequency interactions or beam widening. Therefore, to distinguish these approaches against WBBG, we have used the suffix {\it NBBG} representing a \textit{narrowband beam-gain} approach for these structures. In the numerical results, we considered two setups: mmWave with $f_0 = 28$~GHz and $N = 64$ antennas, and sub-THz with $f_0 = 140$~GHz and $N = 256$ antennas. For the sub-THz, the attenuation by atmospheric gases was modeled according to ITU-R Recommendation P.676 with ambient temperature $T = 15^\circ$C, dry-air pressure $P = 101300$~Pa, and water-vapor density $W = 7.5,\mathrm{g/m^3}$. 
To simplify the analysis, we ignore the bandwidth-dependent performance degradation of TTDs, which is commonly negligible in practice. 
While the Full-TTD-NBBG architecture can, in theory, eliminate beam squint entirely, making its performance nearly independent of the fractional bandwidth, the practical limitations of wideband PSs in Sparse-TTD and Non-TTD architectures introduce performance degradation as the operating bandwidth increases. To quantify this, we consider a simplified performance model.  Consider a ULA transmitter having $N$ antennas with half-wavelength interelement spacing for the central carrier frequency $f_0$, transmitting over $2M+1$ uniformly spaced carrier frequencies across bandwidth $B$.  For simplicity, a single antenna receiver is considered. Let the fractional baseband carrier vector be defined as $\boldsymbol{b} = \left[-\frac{B}{2f_0},\ldots,\frac{B}{2f_0}\right]^{\mathrm{T}}_{(2M+1)\times 1}$, and let $G_{\mathrm{NB}}(b, \theta_{\mathrm{u}})$ denote the array gain at fractional baseband carrier $b$ and observation angle $\theta_{\mathrm{u}}$.

First, we examine the Non-TTD narrowband beamforming performance gap. To do so, we start by formulating the performance of the ideal Full-TTD-NBBG represented as  $R_{\mathrm{FN}}=\sum_{m=-M}^{M} \log\left({1 + \tilde{\gamma}_mN\eta}\right)$ where $\tilde{\gamma}_m$ is the SNR at each antenna element at carrier $m$\footnote{
Note that the Line-of-Sight (LoS) assumption, while simplifying the analysis across the three architectures, also {\bf represents the worst-case analysis} in terms of beam squint mitigation. This is because the multipath propagation resulting from a non-LoS channel assumption provides a high angular diversity, which inherently decreases the amount of beam squint.
}
, $N$ corresponds to the maximized array gain equal to the number of antenna elements, and  $\eta < 1$ captures the performance degradation due to hybrid beamforming relative to a fully digital implementation. In a similar way, if we replace the TTDs by PSs and implement the conventional narrowband beamforming, the performance of the Non-TTD-NBBG can be modeled as $R_{\mathrm{NN}}=\sum_{m=-M}^{M} \log\left(1 + \tilde{\gamma}_mG(b_m, \theta_{\mathrm{u}})\mathcal{E}(b_m)\eta\right)$. Here, two factors reduce performance: (a) $\mathcal{E}(b)<1$, which is a decreasing function of $b$ that models the impact of frequency-dependent hardware impairments on the phase shifters evaluated at $f = f_0(1 + b)$; and (b) $G(b_m, \theta_{\mathrm{u}})\equiv G_{\mathrm{NB}}(b_m, \theta_{\mathrm{u}})$ which represents the narrow-band beam gain for each carrier $b_m$ is lower than the optimal array gain $N$ due to the beam squint. This  can be approximated by the array gain of a fully digital beamformer with ideal PSs, given by $G_{\mathrm{NB}}(b, \theta_{\mathrm{u}}) \approx \left| \frac{\sin(N\pi \Delta_b)}{\sin(\pi \Delta_b)} \right|$, where $\Delta_b = \frac{b}{2} \sin(\theta_{\mathrm{u}})$. Finally, the performance gap between the Full-TTD-NBBG and Non-TTD-NBBG denoted by $\Delta R_{\mathrm{FN-NN}} = R_{\mathrm{FN}} - R_{\mathrm{NN}}$ is represented by 
\begin{align}
\label{eq:perf}
    \Delta R_{\mathrm{FN-NN}}=\sum_{m=-M}^{M} \log\left(\frac{1 + \tilde{\gamma}_mN\eta}{1 + \tilde{\gamma}_mG(b_m, \theta_{\mathrm{u}})\mathcal{E}(b_m)\eta}\right).
\end{align}
The normalized performance gap corresponding to $\Delta R_{\mathrm{FN-NN}}$ is shown in Fig.~\ref{fig:pc-c}-a for the two simulation scenarios at mmWave and THz frequency bands, where a simple linear model is used for the $\mathcal{E}(b)$, with unity gain at $b=0$ and a maximum degradation of 6~dB at $b=\pm \frac{B}{2f_0}$. It is observed that the gap becomes significantly large for $B/f_0 > 0.05$, rendering this architecture suitable only for narrowband scenarios. 

The Non-TTD-WBBG beamformer mitigates this gap by optimizing a wideband beam pattern $G_{\mathrm{WB}}(b_m, \theta_{\mathrm{u}})$ that remains approximately flat across all carriers $b_m$ via a max-min beam gain optimization. This optimization is a complex non-convex problem which can be approximately solved using convexification techniques, such as those in~\cite{8006493}. 
Once the beam pattern is optimized through calculating the optimal value of $G_{\mathrm{WB}}(b_m, \theta_{\mathrm{u}}),\forall m$, the corresponding performance gap between the Full-TTD-NBBG and Non-TTD-WBBG is obtained as $\Delta R_{\mathrm{FN-NW}} = R_{\mathrm{FN}} - R_{\mathrm{NW}}$ obtained from \eqref{eq:perf} where $G(b_m, \theta_{\mathrm{u}})=G_{\mathrm{WB}}(b_m, \theta_{\mathrm{u}})$. 
As depicted in Fig.~\ref{fig:pc-c}-a, the Non-TTD-WBBG architecture substantially reduces the performance gap compared to the Non-TTD-NBBG approach. However, the gap increases with fractional bandwidth due to two main factors: the degradation of wideband beam gain caused by the increased angular coverage required to accommodate squinted beams, and the worsening hardware impairments of PSs at higher fractional bandwidths. 
While this work does not explicitly model all hardware non-idealities to maintain a tractable and insightful comparison of the architectures, various impairments affect each structure to different extents in practice. For instance, phase noise predominantly impacts phase-shifter-based architectures (Non-TTD-NBBG and Sparse-TTD), especially at high fractional bandwidths. Other impairments, such as thermal drift and insertion loss, influence both PSs and TTD units, and therefore affect all three architectures to varying degrees.

Now we discuss the performance of the Sparse-TTD-NBBG beamformer.  Depending on the structure of the hybrid beamformer and the number of RF chains and TTDs employed  for the Sparse-TTD architecture, the performance can be analyzed to lie between two bounds. The lower bound corresponds to the case where no TTDs are used, which can be approximated by the performance of the Non-TTD-WBBG architecture. The upper bound is achieved when the maximum number of TTDs (i.e., $N$) is utilized together with $N$ available PSs in a fully digital beamformer structure, under the assumption that PSs are still present, unlike in the Full-TTD configuration. This corresponds to allocating each antenna element a dedicated TTD cascaded with a PS. Since the cascade of a TTD and a PS is functionally equivalent to a TTD, the performance of this configuration is similar to that of Full-TTD, but with additional degradation due to the bandwidth-dependent impairments of the PSs. Accordingly, the performance upper bound can be approximated as $\sum_{m=-M}^{M} \log\left(1 + \tilde{\gamma}_mN\mathcal{E}(b_m)\right)$. Depending on the design parameters, the actual performance of the Sparse-TTD-NBBG beamformer will lie somewhere between these bounds, as illustrated by the green-shaded region in Fig.~\ref{fig:pc-c}-a.  Furthermore, the performance gap between the sub-THz and mmWave cases is significantly larger for the Non-TTD-NBBG scenario than that related to the Non-TTD-WBBG scenario. The reason is that at sub-THz frequencies, the larger number of array elements yields much sharper narrowband beams, which increases the likelihood of beam misalignment and thus degrades performance (except for the central carrier which experiences no squint). In contrast, in WBBG the beams are deliberately broadened, which mitigates misalignment and reduces the sensitivity of performance to array size differences between the mmWave and sub-THz regimes.

Beyond performance, practical deployment also necessitates consideration of implementation cost. Due to the wide range of parameters influencing cost, the literature lacks a definitive comparison of the cost across different wideband hybrid beamforming architectures. However, estimated comparisons can be made based on qualitative cost considerations and implementation assumptions. Assuming similar baseband and RF chains across all architectures represented in Fig.~\ref{fig:structures}, the primary cost differences arise from the use of TTD units and wideband PSs.  
First, we compare the costs of the Non-TTD-NBBG and Full-TTD-NBBG architectures, represented by the orange and green curves in Fig.~\ref{fig:pc-c}-b. At small bandwidth, the Non-TTD-NBBG is cheaper; however, its cost grows sharply with  bandwidth due to the complexity of wideband phase shifters (multi-section cascaded architecture of wideband PSs \cite{cao2016miniaturized}, complex structures for wideband NGD compensation \cite{meng2022broadband}, and several other practical challenges), eventually surpassing the Full-TTD whose TTDs cost are less sensitive to the bandwidth increase. We now elaborate on why the cost of the Sparse-TTD architecture (blue curve) falls between those of the other architectures for relatively low (below a certain threshold, $\mathrm{th}_1$) and very high (above another threshold, $\mathrm{th}_2$) fractional bandwidths, as observed in the figure. For narrowband operation ($B/f_0<\mathrm{th}_1$), the high cost of TTDs dominates. Consequently, the Full-TTD, Sparse-TTD, and Non-TTD correspond to the highest, intermediate, and lowest costs, respectively, reflecting their use of the largest, moderate, and zero number of TTDs. For ultra-wideband operation ($B/f_0>\mathrm{th}_2$), the escalating cost of wideband phase shifters dominates, reversing the order: Non-TTD becomes most expensive, Full-TTD least, and Sparse-TTD again lies in between.
Among the studied architectures, Non-TTD-NBBG offers the lowest cost but suffers from severe performance degradation at higher fractional bandwidths, limiting its suitability to narrowband systems. The cost trend with respect to the desired fractional bandwidth for the other beamforming architectures is schematically illustrated in Fig.~\ref{fig:pc-c}-b.   Sparse-TTD architecture commonly incurs higher costs than the Non-TTD counterparts due to the additional expense of TTD units. However, this cost gap narrows as the fractional bandwidth increases, since wideband PSs, employed in both Sparse-TTD and Non-TTD-WBBG architectures, become increasingly expensive at higher bandwidths, with their cost growth rate eventually surpassing that of TTDs. 
More specifically, in ultra-wideband scenarios, the cost of Sparse-TTD and Non-TTD-WBBG architectures may ultimately exceed that of the Full-TTD as illustrated in Fig.~\ref{fig:pc-c}-b. This is due to the increased complexity of compensating components (NGDs) 
and calibration mechanisms, required to maintain constant phase for wideband PSs at extreme bandwidths, as discussed in Section~\ref{sec:345kjsdg2}. 

\begin{figure}
    \centering
    \includegraphics[width=1\linewidth]{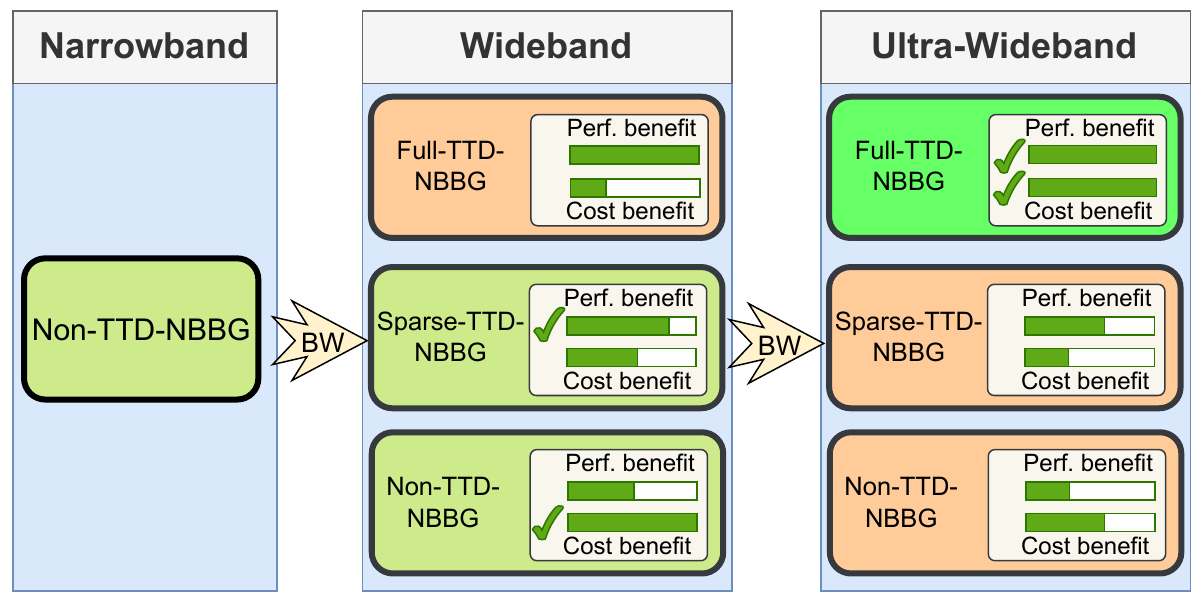}
    \vspace{-15pt}
    \caption{Beam squint mitigation in wideband hybrid beamformers considering the cost-performance trade-off. The normalized performance and cost benefits in each of the wideband and ultra-wideband scenarios are also visualized.
    }
    \label{fig:final-compare}
\end{figure}

Finally, an overall cost-performance analysis can be conducted to guide the selection of the optimal architecture 
as illustrated in Fig.~\ref{fig:final-compare}. Based on the preceding discussion, the Non-TTD-NBBG and Full-TTD-NBBG hybrid beamformers clearly emerge as the preferred choices for narrowband and ultra-wideband scenarios, respectively. More specifically, the Full-TTD-NBBG beamformer is the preferred option in ultra-wideband scenarios in terms of {\bf both performance and cost} as shown in Figs. \ref{fig:pc-c}-a and \ref{fig:pc-c}-b, respectively, and highlighted in  Fig.~\ref{fig:final-compare}.   In contrast, the preferred option for wideband scenarios can be Sparse-TTD-NBBG or Non-TTD-WBBG; however, the decision between the optimal architecture is more nuanced, as it depends on specific design parameters and the relative emphasis placed on performance versus implementation cost. 

\section{Conclusions}
In this work, we presented a comparative evaluation of Non-TTD, Sparse-TTD, and Full-TTD hybrid beamforming architectures for mitigating beam squint in wideband and ultra-wideband communication systems, examining for each the underlying technological principles, implementation challenges, and recent advancements reported in the literature. A key insight from our analysis is the often-overlooked practical limitation imposed by wideband PSs, which significantly affects the cost, complexity, and performance of both Non-TTD and Sparse-TTD architectures, constraining their practical efficiency particularly in ultra-wideband scenarios. Accordingly, our study concludes that Full-TTD-based beamformers potentially offer the most efficient solution for ultra-wideband systems not only in terms of performance but also in terms of cost, since they eliminate the need for wideband PSs.




\bibliographystyle{IEEEtran}
\bibliography{References}
\vspace{-10pt}

\begin{IEEEbiographynophoto}{Mehdi Monemi} 
    \input{bios/bio-monemi-short}
\end{IEEEbiographynophoto}
\vspace{-10pt}

\begin{IEEEbiographynophoto}{Mohammad Amir Fallah} 
    \input{bios/bio-fallah-short}
\end{IEEEbiographynophoto}
\vspace{-10pt}

\begin{IEEEbiographynophoto}{Mehdi Rasti} 
    \input{bios/bio-rasti-short}
\end{IEEEbiographynophoto}
\vspace{-10pt}

\begin{IEEEbiographynophoto}{Omid Yazdani} 
    \input{bios/bio-yazdani-short}
\end{IEEEbiographynophoto}
\vspace{-10pt}

\begin{IEEEbiographynophoto}{Onel L. A. López} 
    \input{bios/bio-onel-short}

\end{IEEEbiographynophoto}
\vspace{-10pt}

\begin{IEEEbiographynophoto}{Matti Latva-aho} 
    \input{bios/bio-matti-short}
\end{IEEEbiographynophoto}

\end{document}

%% file: bios/bio-monemi-short.tex
[M] (mehdi.monemi@oulu.fi) 
is an Adjunct Professor with the Centre for Wireless Communications (CWC), University of Oulu, Finland. His
current research interests include resource allocation in 5G/6G
networks, as well as the employment of AI and ML in wireless networks and verticals. 

%% file: bios/bio-fallah-short.tex
(mfallah@pnu.ac.ir)
is currently an assistant professor with the Department of Engineering, P
ayame Noor University (PNU), Tehran, Iran. His current research interests include antenna and propagation, mobile computing, and the application of machine learning and artificial intelligence in wireless networks.

%% file: bios/bio-rasti-short.tex
[SM] (mehdi.rasti@oulu.fi) 
is an Associate Professor with the Centre for Wireless Communications, University of Oulu, Finland. 
        His current research interests include radio resource allocation in IoT, Beyond 5G and 6G wireless networks, as well as applications of AI in wireless networks and verticals.

%% file: bios/bio-yazdani-short.tex
holds a PhD of wireless communications engineering,
focusing on sustainable resource allocation and connectivity for 5G and 6G networks.

%% file: bios/bio-onel-short.tex
[S’17, M’20, SM’24] (onel.alcarazlopez@oulu.fi) is
an Associate Professor of wireless communications engineering,
focused on sustainable IoT connectivity, at 6G Flagship, University of Oulu, Finland.

%% file: bios/bio-matti-short.tex
[F] (matti.latva-aho@oulu.fi) received Dr.Tech. (Hons.) degree in Electrical Engineering from the University of Oulu, Finland, in 1998. Prof. Latva-aho served as the Director of CWC from 1998 to 2006 and later as Head of the Department of Communication Engineering until August 2014. Currently, he is the Director of the National 6G Flagship Programme and Global Fellow at The University of Tokyo. He has published over 600 conference and journal publications.